# Mobile-Network Connected Drones:
# Field Trials, Simulations, and Design Insights

Xingqin Lin, Richard Wiren, Sebastian Euler, Arvi Sadam, Helka-Liina Maattanen, Siva D. Muruganathan, Shiwei Gao, Y.-P. Eric Wang, Juhani Kauppi, Zhenhua Zou, and Vijaya Yajnanarayana

Ericsson

Contact: xingqin.lin@ericsson.com

**ABSTRACT**

*Drones are becoming increasingly used in a wide variety of industries and services and are delivering profound socioeconomic benefits. Technology needs to be in place to ensure safe operation and management of the growing fleet of drones. Mobile networks have connected tens of billions of devices on the ground in the past decades and are now ready to connect the drones flying in the sky. In this article, we share some of our findings in cellular connectivity for low altitude drones. We first present and analyze field measurement data collected during drone flights in a commercial Long-Term Evolution (LTE) network. We then present simulation results to shed light on the performance of a network when it is serving many drones simultaneously over a wide area. The results, analysis, and design insights presented in this article help enhance the understanding of the applicability and performance of providing mobile connectivity to low altitude drones.*

**INTRODUCTION**

Drones find applications in a wide variety of industries and services [1]. To ensure safe operation and management of the growing fleet of drones, technology needs to be in place to authenticate, monitor, and control the drones, with connectivity support from backbone communication networks for command and control as well as payload communications [2].

Mobile networks are well positioned to assist with drone traffic control and law enforcement [3]. The network solutions specified by the 3rd Generation Partnership Project (3GPP) provide international standard support for mobile networks to offer secure, low latency, and high data rate communication services. Reusing the existing terrestrial mobile networks can facilitate the rapid growth of the drone ecosystem, without the need of developing completely new technologies and big network investments. There may however be challenges related to interference as well as mobility [4]. To better understand the potential of mobile networks for low altitude drones, 3GPP has dedicated a significant effort during its Release 15 to study enhanced Long-Term Evolution (LTE) support for connected drones [5]. Our results presented in this article further sheds light on the applicability and challenges of using LTE networks for providing beyond visual line-of-sight (LOS) connectivity to the drones.

Connected drones have recently drawn much academic interest. However, most of the existing works have been focused on theoretical analysis. Exemplary theoretical works include joint trajectory and communication design [6], three-dimensional coverage analysis [7], spectrum sharing [8], caching [9], and hover time optimization [10]. Channel measurement campaigns have been carried out for aerial channel modeling, see e.g. [11] [12]. By means of measurements and simulations, the work [13] studies the impact of interference and path loss for connected drones. The recent measurement results reported in [14] show that a high number of neighboring cells may be interfered due to uplink transmission from a drone to its serving cell. However, little work has been done to explore mobility, latency, and cell association aspects of mobile-network connected drones.

This article aims to deliver practical, current information, field measurements, state-of-the-art simulation results, and best industry practices on mobile-network connected drones. Specifically, we focus on providing design insights based on measurement campaigns and simulation results. Such insights drawn from experimental measurement campaigns and simulations have high practical relevance. The system design insights discussed can provide guidance to future work in the area. The measurement-campaign-driven study to understand the variations in the received cellular signal statistics by contrasting them with the terrestrial measurements for a mobile drone is of much value. This article also describes how the latency and physical resource allocation for drone connectivity are related. The inferences drawn here are critical for the downlink command-and-control communication for the drones. This article further concretely identifies the main challenges in providing mobility support to the connected drones. Last but not least, this article presents a pioneering contribution that reveals the various cell association patterns in the sky and explains the reasons for the increased radio link failures (RLFs) for the connected drones. This novel discovery may help energize future researchers to work towards mitigating the problems. In summary, the in-depth results and analyses in this article provide novel, valuable design insights that will be of much interest to the audience working in the field.

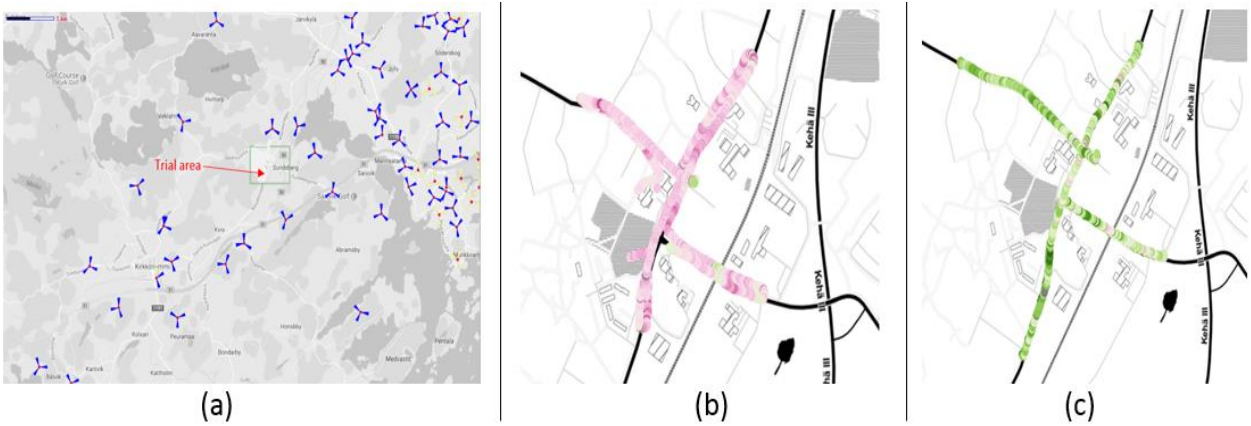

*Figure 1: Drone flight test in Masala, Finland: (a) shows the positions of cell sites and the orientations of sector antennas of BSs around the test area; (b) shows the drone route projected onto a 2D map; (c) shows the reference route of a car running on the ground.*

## FIELD TRIALS

### MEASUREMENT SETUP

In this section, we describe the measurement setup for the field measurement results collected in a commercial LTE network in a suburban area in Masala, Finland. Figure 1(a) shows the positions of cell sites and the orientations of sector antennas of base stations (BSs) around the test area.

Drone flights and measurements were performed by a consumer grade radio-controlled quadcopter DJI Phantom 4 Pro. The maximum flight time of the drone is ~30 minutes. Measurement data were collected with TEMS Pocket 16.3 installed on an LTE smart phone, which was mounted on the drone.

The results and analysis presented in later sections are derived from mobility routes on the ground, at 50 m height, and at 150 m height. The ground level was chosen as the benchmark. The height of 150 m was chosen because that is the maximum permissible drone flight height in Finland where the measurements were conducted. The intermediate height of 50 m was considered as an interesting scenario since this height is close to the BS antenna height in rural areas and using drones for cell site inspection is an emerging important use case.

The results on the ground were collected by driving a car on the ground along the route shown in Figure 1(c). The drone routes in the sky, shown in Figure 1(b), closely follow the route on the ground. The flying speed of the drone was ~18 km/h. The driving speed of the car varied a bit (from 20 km/h to 40 km/h) due to traffic on the ground.

Next, we compare the measurement data collected in the sky to their counterparts collected on the ground. For the sake of brevity, we only present measurement data collected in the 800 MHz band, though we measured other bands as well during the trials.

### SERVING CELL RESULTS AND ANALYSIS

We start by investigating the field measurement data corresponding to the serving cell, as shown in Figure 2.

Reference signal received power (RSRP) is a key measurement parameter indicating the received signal power level in an LTE network. From the RSRP distributions in Figure 2(a), we can see that RSRPs at the heights of both 50 m and 150 m are higher than the RSRPs at the ground level. The $50^{th}$ percentile RSRPs at the heights of 50 m and 150 m are 18.9 dB and 15.6 dB higher than the corresponding value at the ground level, respectively.

The commercial cellular network is equipped with down-tilted BS antennas to optimize terrestrial coverage. At the height of 50 m or 150 m, the drone user equipment (UE) was likely served by either the low gain parts of the main lobes or the sidelobes of the BS antennas, which have reduced antenna gains compared to the main lobes serving UE on the ground. The propagation conditions in the sky, however, are close to free-space propagation. As a result, we observed that the drone UE experienced stronger serving cell RSRPs than ground UEs.

Reference signal received quality (RSRQ) is another key measurement parameter indicating the received signal quality level in an LTE network. RSRQ includes the effect of interference from neighbor cells. From the RSRQ distributions in Figure 2(b), we can see that RSRQs at the heights of both 50 m and 150 m are lower than the RSRQs at the ground level. The $50^{th}$ percentile RSRQs at the heights of 50 m and 150 m are 2.8 dB and 4.1 dB lower than the corresponding value at the ground level, respectively. Worse serving cell RSRQs in the sky are expected due to the close to free-space propagation which leads to stronger downlink interference from non-serving cells to the drone UE.

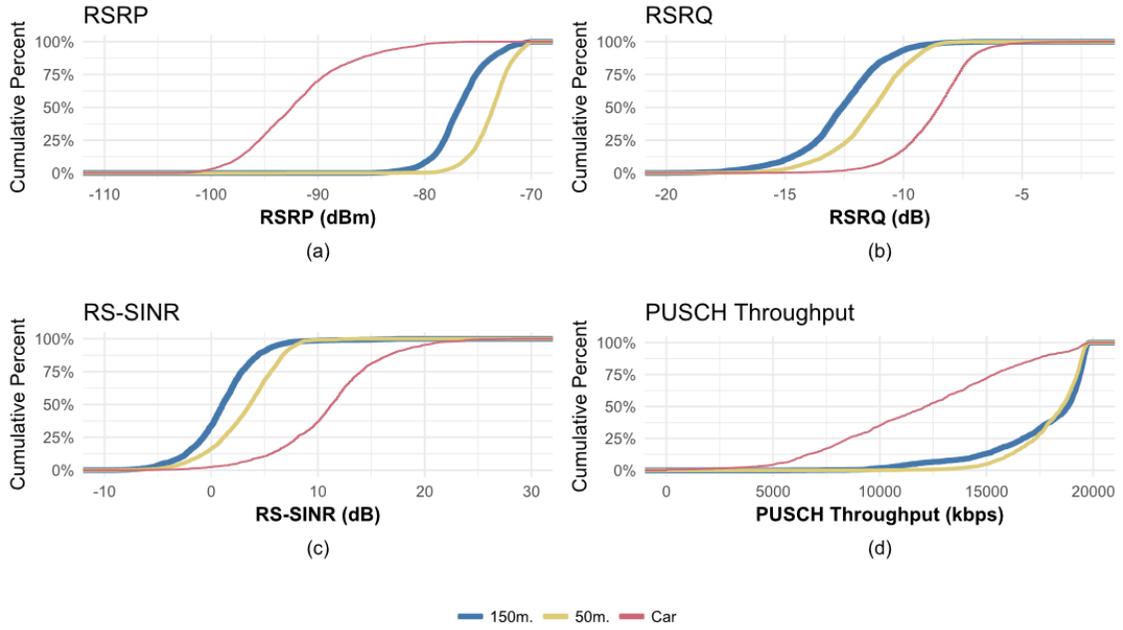

*Figure 2: Serving cell measurement data: (a) shows distributions of RSRP; (b) shows distributions of RSRQ; (c) shows distributions of RS-SINR; (d) shows distributions of uplink throughput.*

An alternative received signal quality metric is signal to interference-plus-noise ratio (SINR). Figure 2(c) shows the distributions of reference signal SINR (RS-SINR). The $50^{th}$ percentile RS-SINRs at the heights of 50 m and 150 m are 7.7 dB and 10.2 dB lower than the corresponding value at the ground level, respectively. These RS-SINR results indicate again the much stronger downlink interference from non-serving cells in the sky than at the ground level.

The received signal power at the drone UE from a BS mainly depends on two factors: pathloss and antenna gain. In general, as the height increases, the pathloss first decreases due to the increased LOS probability and then increases because the loss due to increased link distance becomes higher than the further marginal gain from increased LOS probability. Additionally, the antenna gain of the sidelobe decreases as the height increases. As a result, as the height increases, the received signal power at the drone UE from the BS first increases and then decreases, as shown by the field results in Figure 2(a). The downlink interference power is a sum of the received signal powers from all the neighboring BSs. In general, as the height increases, the downlink interference power also first increases and then decreases. Similar reasoning applies to the uplink interference as well. For the height range considered in this article (maximum height of 150 m in the field trials and maximum height of 300 m in the simulations), we observe that the signal quality (SINR/RSRQ) decreases as the height increases.

Note that RSRP, RSRQ, and RS-SINR are all downlink metrics. Many drone use cases such as flying cameras and remote surveillance are likely to be uplink data (from drone UE to BS) heavy. In the field measurements, we also logged the uplink throughputs over LTE physical uplink shared channel (PUSCH) associated with file uploading, as presented in Figure 2(d). We can see that the variances of uplink throughputs at the heights of both 50 m and 150 m are much smaller than at the ground level. We also observe that throughputs at the heights of both 50 m and 150 m are higher than the corresponding values at the ground level. The $50^{th}$ percentile throughput at the heights of 50 m and 150 m are ~18.5 Mbps and ~18.8 Mbps, respectively, while the counterpart throughput at the ground level is ~12.5 Mbps. In this field trial, the drone UE likely experienced better uplink channel conditions because of close to free-space propagation conditions in the sky when compared to the propagation conditions on the ground. This likely resulted in better uplink throughputs in the sky. It should however be noted that the throughput performance heavily depends on many factors such as network load and scheduling.

## NEIGHBOR CELL RESULTS AND ANALYSIS

In this section, we examine the field measurement data corresponding to the neighbor cells. The analysis of neighbor cell signal strengths and quality values is useful since it sheds light on the impact of interference from neighboring cells. In addition, neighbor cell signal strengths and quality values are important factors for mobility procedures. In a typical network, UE is configured to measure and report neighbor cell signal

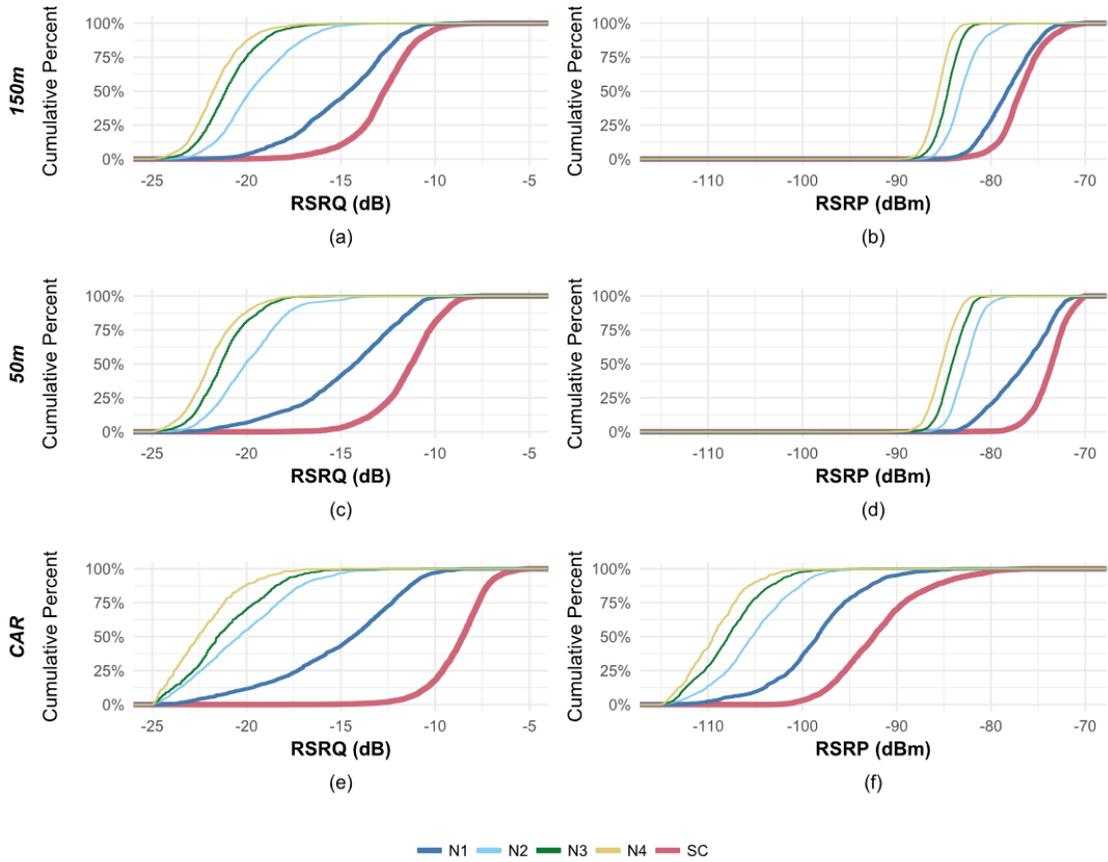

*Figure 3: Neighbor cells measurement data: (a), (c), and (e) show distributions of RSRQ at the heights of 150 m and 50 m, and at the ground level; (b), (d), and (f) show distributions of RSRP at the heights of 150 m and 50 m, and at the ground level. Nk denotes the k-th strongest neighbor cell, and SC denotes the serving cell.*

strengths and quality values which will be used by the network to make handover decisions.

The distributions of RSRP and RSRQ data for the four strongest neighbor cells are shown in Figure 3. We use N$k$ to denote the $k$-th strongest neighbor cell. For ease of comparison, we also plot the distributions of the serving cell RSRP and RSRQ. The first row of the plots shows the distributions of RSRP and RSRQ data at the height of 150 m, the second row at the height of 50 m, and the third row at the ground level.

From the subfigures (b), (d), and (f) of Figure 3, we can see that the neighbor cell RSRP spread decreases as the height increases. The 50[th] percentile RSRP difference between the first (N1) and the fourth (N4) strongest neighbor cell decreases from 11.4 dB at the ground level to 7.4 dB at the height of 50 m and to 6.5 dB at the height of 150 m. For the neighbor cells other than the strongest, the 50[th] percentile RSRPs are below -105 dBm at the ground level. At the height of either 50 m or 150 m, since the propagation conditions to the drone UE are close to free-space propagation with mostly LOS, the 50[th] percentile RSRPs of these neighbor cells are above -86 dBm.

From the subfigures (a), (c), and (e) of Figure 3, we can observe similar trends for RSRQ.

From the UE's perspective, the relative received signal strengths between the detected neighbor cells and the serving cell are more relevant. Note that the values of these RSRP gaps (serving cell RSRP - neighbor cell RSRP) can be negative since the serving cell in practice is not always the strongest cell due to the dynamic network environment and handover margin used. For the RSRP gap between the serving cell and the strongest neighbor cell (serving cell RSRP - N1 RSRP), the percentage of negative RSRP gaps at the ground level is ~11% in the field measurement. The percentage of negative RSRP gaps increases as the height increases to ~21% at 50 m and ~33% at 150 m. This is likely due to faster antenna pattern roll-off as the drone UE is more often served by either the sidelobes or the low gain part of the main lobe of a BS antenna pattern.

In this measurement, we find that the 50[th] percentile RSRP gap between the serving cell and the strongest neighbor cell at the ground level is 6.5 dB. The corresponding RSRP gaps at the heights 50 m and 150 m are 2.8 dB and 1.6 dB, respectively. Typically, a measurement report is triggered when the neighbor cell becomes X dB better than the serving cell, where X is usually set to a small value (e.g. 3 dB). To better control the measurement reports from drone UEs, 3GPP has enhanced measurement reporting in Release-15 LTE such that network can configure measurement report to be triggered when measured RSRPs/RSRQs/RS-SINRs of multiple cells are above a threshold.

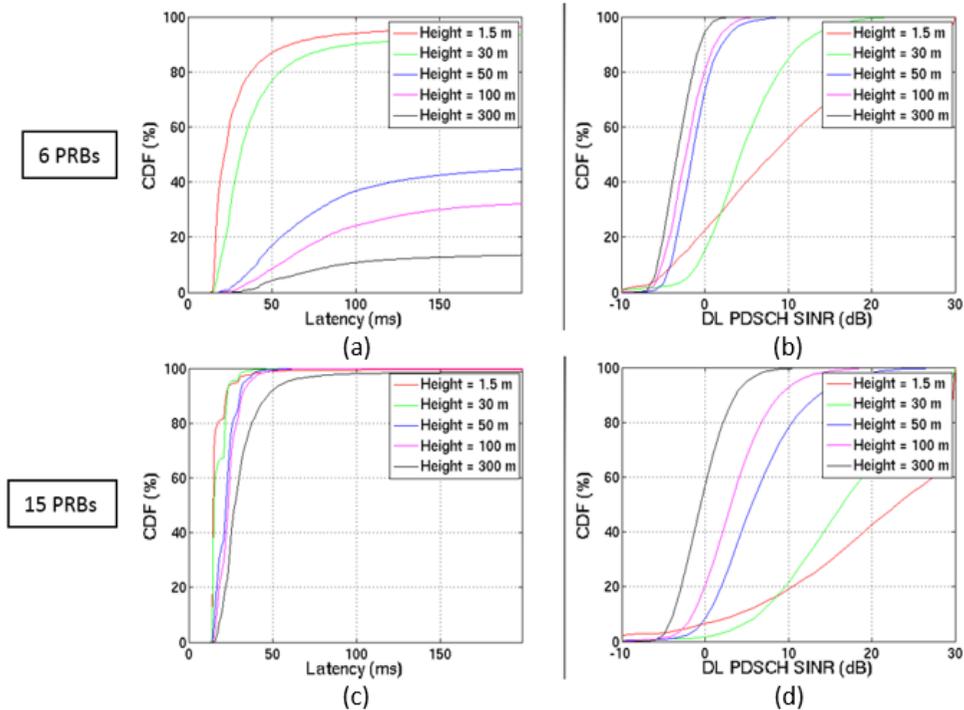

*Figure 4: Downlink latency simulation results: (a) and (c) show latency distributions when 6 and 15 PRBs are used to serve drone UEs, respectively; (b) and (d) show PDSCH SINR distributions when 6 and 15 PRBs are used to serve drone UEs, respectively.*

Field trials are valuable, but they are also limited in terms of number of drones that can be simultaneously tested and the range of features and configurations in an operational commercial network. To complement field trials, we present simulation results in the next two sections to gain insights into the network performance when the network is serving many drone UEs simultaneously over a wide area.

## LATENCY SIMULATION RESULTS AND ANALYSIS

The ability of remote command and control can significantly enhance safety and operation of drones. Such command-and-control communications need to be reliable, and the packets should be successfully delivered within some latency bound (depending on the use case) with high probability. In this section, we present simulation results of radio interface latency for LTE networks serving command-and-control traffic for drones.

*SIMULATION SETUP*

The latency simulation assumptions follow the 3GPP study item on enhanced LTE support for aerial vehicles [5]. For ease of reference, we summarize a few key evaluation assumptions.

- **Traffic model:** packets arrive periodically with a period of 100 ms and have a fixed packet size of 1250 bytes, which leads to a data rate of 100 kbps for command and control at each drone UE. The required latency bound is 50 ms.
- **Deployment scenario:** an urban macro scenario with aerial vehicles, where sites are placed on a hexagonal grid with 19 sites and 3 cells per site. The LTE system bandwidth is 10 MHz at 2 GHz carrier frequency.
- **Antenna model and configuration:** Each BS has two cross polarized TX/RX antennas with 10 degrees of down-tilt at the height of 25 m. The BS antennas are modeled by a synthesized antenna pattern using an antenna array with a column of 16 cross-polarized antenna elements, where the antenna spacing is $0.8\lambda$ where $\lambda$ denotes the wavelength. Each UE has one omni-directional TX and two cross-polarized omni-directional RXs.

In this evaluation, we focus on command-and-control traffic in the downlink and assume that the scheduler partitions the radio resources so that the drone traffic and terrestrial mobile traffic are scheduled in orthogonal frequency resources. With this partition, the signals to terrestrial UEs and the signals to drone UEs do not interfere. However, drone UEs in a cell still experience interference from neighbor cells since the neighbor cells may use the same radio resource to serve other drone UEs connected to the neighbor cells.

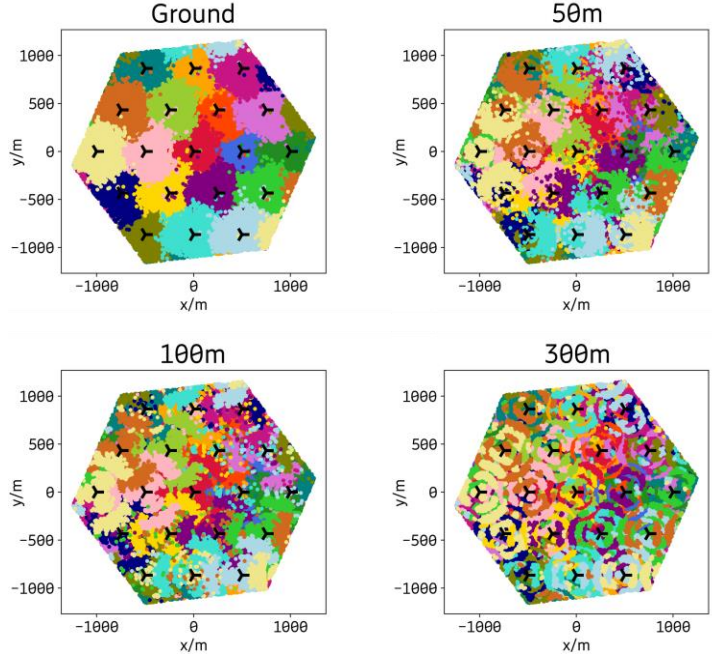

*Figure 5: Maximum-received-power-based cell association patterns at the ground level and at the heights of 50 m, 100 m, and 300 m: UEs in the areas marked by the same colors are associated with the same site.*

We assume 5 drone UEs in each cell, resulting in 500 kbps command-and-control traffic demand per cell. In the evaluation, we simulated the performance at different heights: 1.5 m, 30 m, 50 m, 100 m, and 300 m. We present simulation results below under 2 different numbers of physical resource blocks (PRBs) allocated to the drone UEs: 6 PRBs (i.e., 1.08 MHz) and 15 PRBs (i.e., 2.7 MHz).

## CONNECTED DRONES ALLOCATED WITH 6 PRBS

Figure 4(a) shows the latency distributions at the different heights when 6 PRBs are used to serve the drone traffic. Figure 4(b) shows the corresponding distributions of physical downlink shared channel (PDSCH) SINR.

From the PDSCH SINR distributions, we can see that the SINR drops significantly as the height increases, especially for the heights equal to or higher than 50 m. From the latency distributions, we can see that even at the ground level of 1.5 m, it is not possible to meet the 50 ms latency bound with a high confidence level (e.g. 90%).

To better understand the results, we next examine the resource utilization ratios. The resource utilization ratio is defined as the fraction of utilized radio resources averaged over time, frequency, and cells. It is a key performance indicator that can reflect the interference level in the network.

When 6 PRBs are used to serve the drone traffic, we find that the resource utilization ratios are ~41%, ~57%, ~90%, ~95%, and ~96% at the heights of 1.5 m, 30 m, 50 m, 100 m, and 300 m, respectively. These resource utilization ratios help to explain the results. At the ground level of 1.5 m, the resource utilization is already ~41%, which implies that close to half of the BSs are transmitting and a typical drone UE receiving a downlink packet experiences interference from the corresponding active neighbor cells. As the height increases to 50 m, the resource utilization becomes ~90%, and increases to ~95% as the height increases further. In other words, almost all the BSs are transmitting and a typical drone UE at the height of 50 m or above experiences strong inter-cell interference. The increased resource utilization in the sky is due to poor geometry which in turn leads to lower spectral efficiency because of interference.

## CONNECTED DRONES ALLOCATED WITH 15 PRBS

Figure 4(c) shows the latency distributions at the different heights when 15 PRBs are used to serve the drone traffic. Figure 4(d) shows the corresponding distributions of PDSCH SINR.

From the PDSCH SINR distributions, we can see that though the SINR drops significantly as the height increases, it is much higher than in the case when only 6 PRBs are used to serve the drone traffic. From the latency distributions, we can see that it is possible to meet the 50 ms latency bound with a high confidence level of ~99% at the heights of 1.5 m, 30 m, 50 m, and 100 m, and ~92% at the height of 300 m.

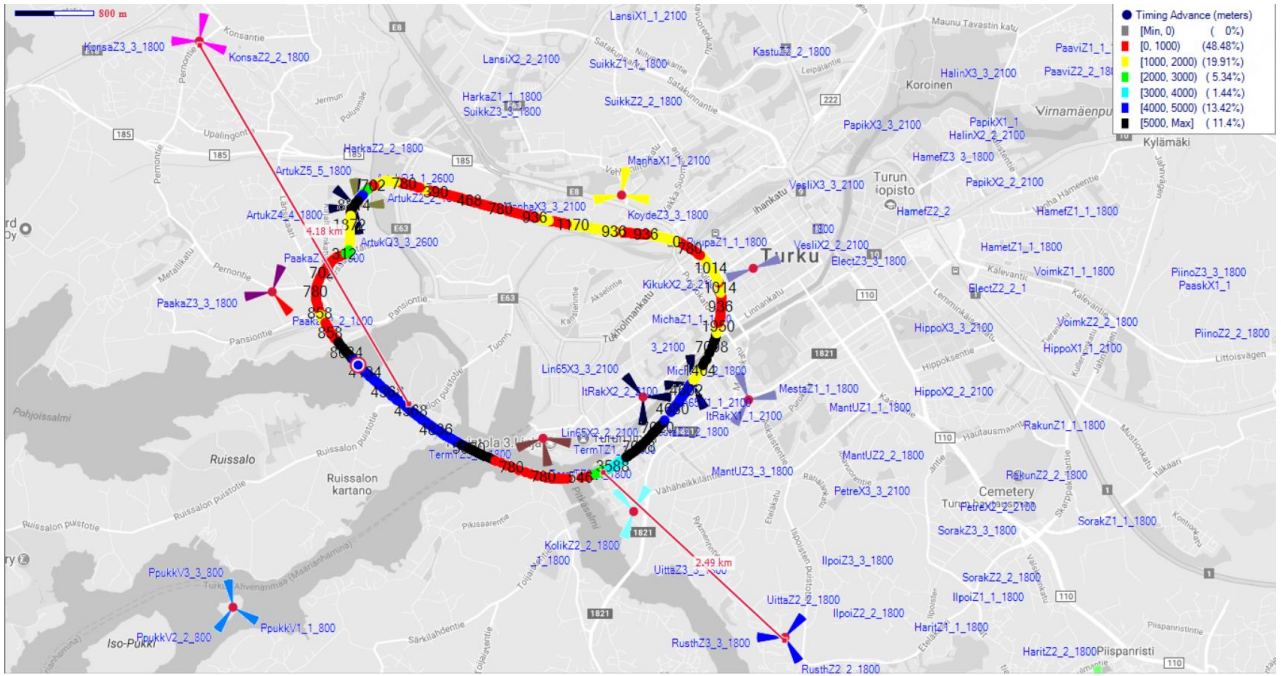

*Figure 6: Cell association for UE flying above 300 m height with a speed of >200 km/h measured in Turku area, Finland.*

When 15 PRBs are used to serve the drone traffic, we find that the resource utilization ratios are ~11%, ~11%, ~23%, ~30%, and ~47% at the heights of 1.5 m, 30 m, 50 m, 100 m, and 300 m, respectively. These resource utilization ratios help explain the results. At the height of 1.5 m, 30 m, 50 m, or 100 m, the resource utilization is not larger than ~30%, and thus the downlink interference experienced at drone UEs is moderate. At the height of 300 m, the resource utilization is ~47%, and thus the downlink interference is stronger.

*REMARKS ON THE LATENCY SIMULATION RESULTS*

The above latency simulation results show the tradeoff between latency performance and the number of PRBs used for the drone command-and-control traffic. Note that the results are under a high command-and-control traffic demand. In the initial phase of drone deployment, it is likely that the demand of command-and-control traffic is much lower.

A general trend we observe from the resource utilization ratios is that as the height increases from 30 m to 300 m, the resource utilization ratio increases for the same offered command-and-control traffic. Height information is thus helpful for the network to perform radio resource management to offer the right service optimization for drone UEs. This has motivated 3GPP to introduce height-based measurement reporting in Release-15 LTE. A key lesson from the above results is that when the resource utilization ratio is low, the downlink interference experienced at drone UEs is not strong, which makes it possible to deliver a small data packet within the 50 ms latency bound with a high confidence level. It is expected that as long as an interference mitigation technique can lead to satisfactory SINR values, it is possible to deliver a small command-and-control packet within some latency bound with a high confidence level.

The latency simulations assume a network with 57 cells and each cell has 5 drone UEs. We do not have the capability to perform such a large field measurement campaign to validate the simulation results. We however find that the simulation results presented in 3GPP Tdoc R1-1720571 [15] well match our results, and thus help validate the accuracy of our latency simulation results.

**MOBILITY SIMULATION RESULTS AND ANALYSIS**

Ensuring reliable connections in the presence of drone movements is important for many drone use cases. There are two main aspects that make mobility support for drone UEs challenging. First is the serving cell signal. BS antennas are typically tilted downwards by a few degrees. The main lobe of a BS antenna thus covers a large part of the surface area of the cell to improve performance for terrestrial UEs. Accordingly, at the ground level the strongest site is typically the closest one. A drone UE on the other hand may be frequently served by the sidelobes of BS antennas, which have lower antenna gains. The coverage areas of the sidelobes may be small and the signals at the edges may drop sharply due to deep antenna nulls. At a given location, the strongest signal might come from a faraway BS, if the gain of the sidelobes of the closer BSs to the drone UE is much weaker. These effects can be clearly seen in Figure 5, which shows the maximum-received-power-based cell association patterns at the ground level and at the heights of 50 m, 100 m, and 300 m. At the higher heights, the coverage areas become fragmented and the fragmentation pattern is determined by the lobe structures of the BS antennas. To further validate the cell association pattern at higher heights, we conducted measurement with a helicopter flying above 300 m in Turku (since the drone height limit is set by the regulators to 150 m in Finland). The cell association pattern from the field measurement is shown in Figure 6. The measurement result further validates the simulation finding that a UE high in the sky may not connect to the closest cells. The red lines in Figure 6 show examples where the UE is connected to cells faraway.

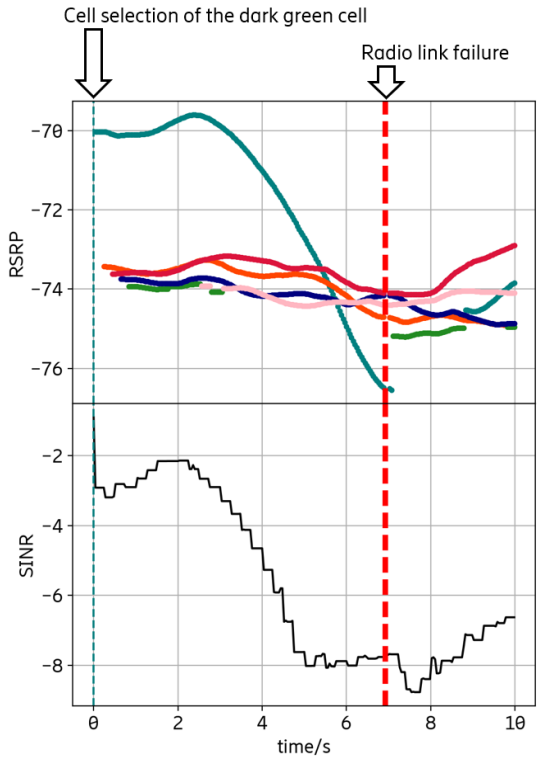

*Figure 7: An example mobility trace for a drone UE moving away from the coverage of a BS antenna sidelobe at the speed of 30 km/h and at the height of 300 m: Each colored RSRP trace corresponds to one cell.*

The second aspect is interference. As pointed out previously, as the height increases, more BSs have LOS propagation conditions to drone UEs. As a result, the drone UEs may generate more uplink interference to the neighbor cells while experiencing more downlink interference from the neighbor cells. Due to the increased interference, SINR could become quite poor at certain heights. The degraded SINR might lead to more RLFs. It might also result in more handover failures since measurement reports, handover commands, etc., may get lost during the handover execution procedure.

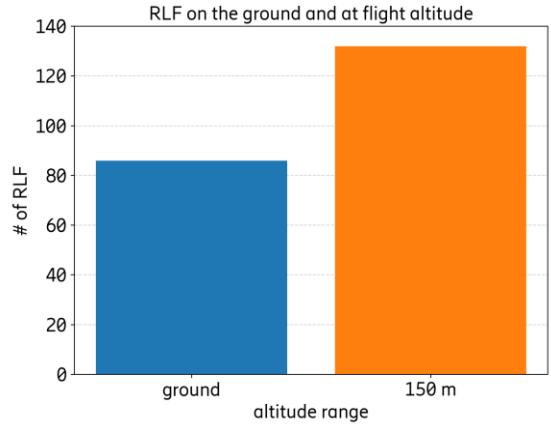

*Figure 8: Number of RLFs in the field trial measured on the ground and at a flight altitude of 150 m.*

To illustrate the challenges of mobility support for drone UEs, Figure 7 shows a simulated example mobility trace for a drone UE moving away from the coverage of a BS antenna sidelobe at the speed of 30 km/h and at the height of 300 m. The upper subfigure of Figure 7 shows the RSRP measurements by the drone UE, and the bottom subfigure shows the time varying trace of the serving cell SINR. Each colored RSRP trace corresponds to a different cell. In Figure 7, the vertical dark green dashed line at the beginning marks cell selection of the dark green cell. At about 3 s, the serving cell RSRP begins to drop, and it drops by 7 dB within 4 s. After 5 s, the neighboring cells become stronger than the serving cell. However, to trigger handover measurement reports, some neighbor cell RSRP should be X dB better than the serving cell, where X is set to 3 in the simulation. From the RSRP traces, we can see that the RSRPs of the neighbor cells stay relatively low, and none of them is at least 3 dB better than the serving cell before the drone UE declares RLF at $t = 7$ s (marked by the vertical red dashed line) due to poor serving cell SINR.

The increased RLF rate at a higher height was also observed in the field measurements. Figure 8 shows the number of RLFs in the field trial measured on the ground and at a flight altitude of 150 m. We can see that the number of RLFs at the height of 150 m increases by more than 50% compared to the ground. The mobility simulation results are in line with the observations in the field trial shown in Figure 8.

To sum up, Figure 7 illustrates the two main aspects that make mobility support for flying drone UEs challenging and interesting: serving cell signal and interference. Since drone UEs may be served by the sidelobes of BS antennas, they might experience sudden drops in signal quality due to antenna nulls when they move from the coverage area of one sidelobe to the coverage area of another sidelobe. The default handover procedure may become too slow to be successfully executed. Further, we can see from Figure 7 that the gaps between the serving cell RSRP and the neighbor cell RSRPs are small, and that the strong interference from neighbor cells makes the serving cell SINR stay relatively low throughout. These observations are consistent with the field measurement results shown in Figure 3 (RSRP) and Figure 2 (SINR). If supplemental data such as flight routes are available to the network, such data can be utilized to facilitate more robust mobility management for drone UEs. In Release-15 LTE, 3GPP has introduced signaling support to request flight path information from UE.

**CONCLUSIONS AND FUTURE WORK**

Mobile networks have connected tens of billions of devices on the ground in the past decades and are now ready to connect the drones flying in the sky. The field measurements presented in this article were conducted in a real commercial LTE network, and the measurement results demonstrate the applicability of terrestrial networks for connected drones. The simulations were conducted fully based on the latest developments in 3GPP including channel models, traffic models, antenna models, and network scenarios. Based on the measurements and simulations, we find that the existing mobile LTE networks targeting terrestrial usage can support the initial deployment of low altitude drones, but there may be challenges related to interference as well as mobility. Future work from other researchers that can help further validate our findings and overcome the identified challenges is desirable.

Providing cellular connectivity to drones is an emerging field. We conclude by pointing out some fruitful avenues for future research.

**Higher-altitude drones:** In this article, we have focused on low-altitude drones with heights up to 300 m. One important extension is to explore the potential of mobile network connectivity for higher-altitude drones. At higher heights, vertical beamforming or up-titled BS antennas may be needed to provide better coverage. The new scenarios will require further analysis, simulations, and field measurements.

**Air-ground channel characterization:** The characteristics of air-ground wireless channels are different from those of terrestrial wireless channels. This is one of the root causes that results in interference and mobility challenges identified in this article. More empirical measurements will be of high value for developing more accurate statistical air-ground channel models. Take Doppler effects for example. In this article, they are implicitly captured in the results, i.e., the measurement and simulation results depend on the speeds of the drones. Characterizing Doppler effects explicitly in the channel measurement campaigns will be of interest, especially for drones flying with high speeds.

**Drones as BSs:** The focus of this article is providing connectivity to the sky, where drones are UEs. There are also ongoing thrusts on providing connectivity from the sky, i.e., drones are BSs. BSs mounted on drones can boost network performance during large events and provide network connections for prompt disaster response. How to integrate a supplemental drone-based network with an existing mobile network deserves further analysis, simulations, and field measurements.

nodes," *IEEE Communications Magazine*, vol. 54, no. 5, pp. 44-50, May 2016.

[14] R. Amorim *et al.*, "Measured uplink interference caused by aerial vehicles in LTE cellular networks," *IEEE Wireless Communications Letters*, to appear.

[15] ZTE, Sanechips, "Evaluation on reliability for LTE aerials," 3GPP Tdoc R1-1720571, November 2017.

## BIOGRAPHIES

**Xingqin Lin** is a Senior Researcher and Standardization Delegate at Ericsson. He leads 4G/5G research and standardization in the areas of drones and satellites. He is a key contributor to 5G NR, NB-IoT, and LTE-M standards specifications. His pioneering work on cellular connected drones helped establish the 3GPP Rel-15 work on enhanced LTE support for aerial vehicles. He served as an editor of the IEEE COMMUNICATIONS LETTERS from 2015-2018. He holds a Ph.D. in electrical and computer engineering from The University of Texas at Austin, USA.

**Richard Wiren** is a Senior Solution Architect at Ericsson and has been working with design, rollout, tuning and optimization of 2G, 3G, 4G and 5G mobile networks since 2000. He has been on several engagements within Industry & Society and IoT, including non-Telco customers looking for new opportunities in utilizing latest mobile network technologies. He is officially registered professional RPAS pilot both in Finland and Sweden where he has conducted UAV operations.

**Sebastian Euler** joined Ericsson Research in 2016. He has since focused on extending the LTE and 5G New Radio (NR) standards with better support for aerial vehicles. In addition, his research interest includes satellite communication networks. He received his PhD in particle physics from RWTH Aachen University in 2014. Afterwards he held a Postdoc position at Uppsala University, and has worked with neutrino experiments in Antarctica during that time.

**Arvi Sadam** received his M.S. degree in Aircraft Engineering from Estonian Aviation Academy, Tartu, Estonia in 2011. He is currently an Experienced Radio Engineer at Ericsson, and has been working with design, rollout, tuning and optimization of 2G, 3G, 4G and 5G mobile networks since 2007. His latest focus is on 3D radio design and creation of spatial radio environment input for UAV missions. He is also a holder of EASA CPL(H) pilot license.

**Helka-Liina Määttänen** received her Ph.D. degree in communications engineering from Helsinki University of Technology in 2012. She joined Ericsson in 2014 and worked before that at Broadcom and Renesas Mobile/Nokia. Her research interest includes downlink MIMO systems, CoMP, LTE-WLAN interworking, FeMBMS, 5G mobility, satellites, IAB and Aerials. She has been attending 3GPP WG1 and is currently attending 3GPP WG2 as a standardization delegate. She was a Rapporteur of LTE Rel-15 WI on Aerials.

**Siva D. Muruganathan** received his PhD degree in electrical engineering from the University of Calgary, Canada in 2008. He is currently with Ericsson Canada working as a Researcher and 3GPP RAN1 delegate. He previously held research/postdoctoral positions at BlackBerry Limited, CRC Canada, and the University of Alberta, Canada. His recent standardization work has been in the areas of MIMO and air-to-ground communications. He was a Rapporteur for the 3GPP Release-15 study on LTE Aerials.

**Shiwei Gao** is a researcher at Ericsson Canada, working on multi-antenna technologies and their standardization in LTE and NR since 2014. He was previously with Nortel Networks from 2007 to 2009 and Blackberry from 2009 to 2013, working on advanced wireless technology research and base station designs. His current research interests include MIMO and their application in wireless communications.

**Y.-P. Eric Wang** is a Principal Researcher at Ericsson Research. He holds a PhD degree in electrical engineering from the University of Michigan, Ann Arbor. Dr. Wang was an Associate Editor of the *IEEE Transactions on Vehicular Technology* from 2003 to 2007. He was a co-recipient of Ericsson's Inventors of the Year award in 2006. Dr. Wang has contributed to more than 150 U.S. patents and more than 50 IEEE articles.

**Juhani Kauppi** has more than 30 years of experience from the telecom industry in which he has always been working with the latest products and solutions in the next generation mobile networks. His current focus areas are in 5G, IoT, drones and implementing data science for analyzing mainly the radio air interfaces.

**Zhenhua Zou** is a researcher at Ericsson, working on ultra-reliable low latency communication and air-to-ground. He received his Ph.D. degree in electrical engineering from KTH Royal Institute of Technology, Sweden. His research interest includes control, optimization, and learning algorithms applied in wireless communication and communication networks.

**Vijaya Yajnanarayana** received the MS degree in electrical engineering (EE) from the Illinois Institute of Technology, Chicago, and PhD degree in EE from the KTH Royal Institute of Technology, Stockholm, Sweden, in 2007 and 2017, respectively. He is a recipient of Program of Excellence Award from KTH Royal Institute of Technology which carried an award of 1 Million Swedish kroner. Currently, he is working as a Researcher in Ericsson Radio Research Group in Stockholm, Sweden.